\documentclass{article}
\usepackage{spconf,amsmath,graphicx}
\usepackage{multirow}
\usepackage{diagbox} 
\usepackage{tabularx}
\usepackage{threeparttable}
\usepackage{graphicx}
\usepackage{amsfonts}
\usepackage{makecell}

\title{Deep Implicit Distribution Alignment Networks for Cross-Corpus Speech Emotion Recognition}
\name{Yan Zhao, Jincen Wang, Yuan Zong$^*$\thanks{* indicates the corresponding authors.}, Wenming Zheng$^*$, Hailun Lian, and Li Zhao}
\address{Key Laboratory of Child Development and Learning Science of Ministry of Education, \\Southeast University, Nanjing 210096, China\\
\{zhaoyan, 220222338, xhzongyuan, wenming\_zheng, lianhailun, zhaoli\}@seu.edu.cn
}

\begin{document}
%
\maketitle
\begin{abstract}
In this paper, we propose a novel deep transfer learning method called deep implicit distribution alignment networks (DIDAN) to deal with cross-corpus speech emotion recognition (SER) problem, in which the labeled training (source) and unlabeled testing (target) speech signals come from different corpora. Specifically, DIDAN first adopts a simple deep regression network consisting of a set of convolutional and fully connected layers to directly regress the source speech spectrums into the emotional labels such that the proposed DIDAN can own the emotion discriminative ability. Then, such ability is transferred to be also applicable to the target speech samples regardless of corpus variance by resorting to a well-designed regularization term called implicit distribution alignment (IDA). Unlike widely-used maximum mean discrepancy (MMD) and its variants, the proposed IDA absorbs the idea of sample reconstruction to implicitly align the distribution gap, which enables DIDAN to learn both emotion discriminative and corpus invariant features from speech spectrums. To evaluate the proposed DIDAN, extensive cross-corpus SER experiments on widely-used speech emotion corpora are carried out. Experimental results show that the proposed DIDAN can outperform lots of recent state-of-the-art methods in coping with the cross-corpus SER tasks.
\end{abstract}

\section{Introduction}

As a typical task of affective computing and speech signal processing, research of SER seeks to empower the computers to automatically understand the emotional states, e.g., \textit {Happy}, \textit{Fear}, and \textit{Disgust}, from the speech signals. It has constantly been under the spotlight over past several decades~\cite{Ayadi2011Survey,schuller2018speech} and lots of promising SER methods have been proposed~\cite{zong2016double,mirsamadi2017automatic,lu2022domain}. However, it is noted that numerous interference factors, \emph{e.g.}, language gap, speaker difference, and corpus variance between the training and testing speech signals, still hinder the possibility of existing well-performing SER methods to move from the laboratory to the practical scenes. This is because that these interference factors would leads to a feature distribution mismatch between the training and testing speech signals and hence remarkably degrade the performance of most well-performing SER methods. To overcome this shortcoming, in recent years some researchers have drawn their attention to a more challenging but fascinating SER issue, \emph{a.k.a.}, cross-corpus SER~\cite{schuller2010cross}. Different from the conventional SER, the labeled training and unlabeled testing speech signals in cross-corpus SER belong to different speech emotion corpora. We also refer the training and testing samples/corpora/features/signals as the source and target ones, respectively.

The earliest contribution to cross-corpus SER can be traced to the work of~\cite{schuller2010cross}, in which Schuller \emph{et al.} proposed a series of feature normalization methods including corpus normalization, speaker normalization, and corpus-speaker normalization to eliminate the corpus difference between the source and target speech signals. Subsequently, several researchers tried to treat cross-corpus SER as a transfer learning task and proposed lots of well-performing transfer subspace learning and deep transfer learning methods. For example, Liu \emph{et al.}~\cite{liu2018unsupervised} proposed a novel transfer subspace learning method called domain-adaptive subspace learning (DoSL) to learn a common subspace to remove the feature distribution mismatch between the source and target speech samples by minimizing their marginal MMD. In the work of~\cite{zhao2022deep}, Zhao \emph{et al.} proposed a novel deep regression method called deep transductive transfer regression networks (DTTRN), whose major module designed for adapting source and target speech feature distributions are still based on the variant of MMD, \emph{i.e.}, multi-kernel MMD. Besides MMD based methods, adversarial learning is also widely-used in dealing with the cross-corpus SER problem~\cite{abdelwahab2018domain,su2022unsupervised}. Unlike MMD and its variants, these methods are not straightforward ones. This means they align the feature distribution gap using an implicit way, \emph{i.e.}, leveraging a domain (corpus) discriminator to disable the networks to be aware of corpus variance.


Inspired by the success of the adversarial learning based methods, in this paper we also focus on the research of cross-corpus SER from the angle of implicit distribution alignment (IDA) and propose a novel deep transfer learning method called deep implicit distribution aligned networks (DIDAN). The major contribution of the proposed DIDAN is designing a novel IDA term to calibrate the feature distribution gap caused by the speech corpus variance. Moreover, different from both MMD and adversarial learning based methods, our IDA performs corpus invariant feature learning by enforcing the learned target speech features to be sparsely reconstructed by the source ones. Hence, all target samples gradually become more involved in the source speech corpus and eventually share the same or similar feature distribution with the source ones.

\section{Proposed method}

\begin{figure}[t]
\centering
\includegraphics[width=\columnwidth]{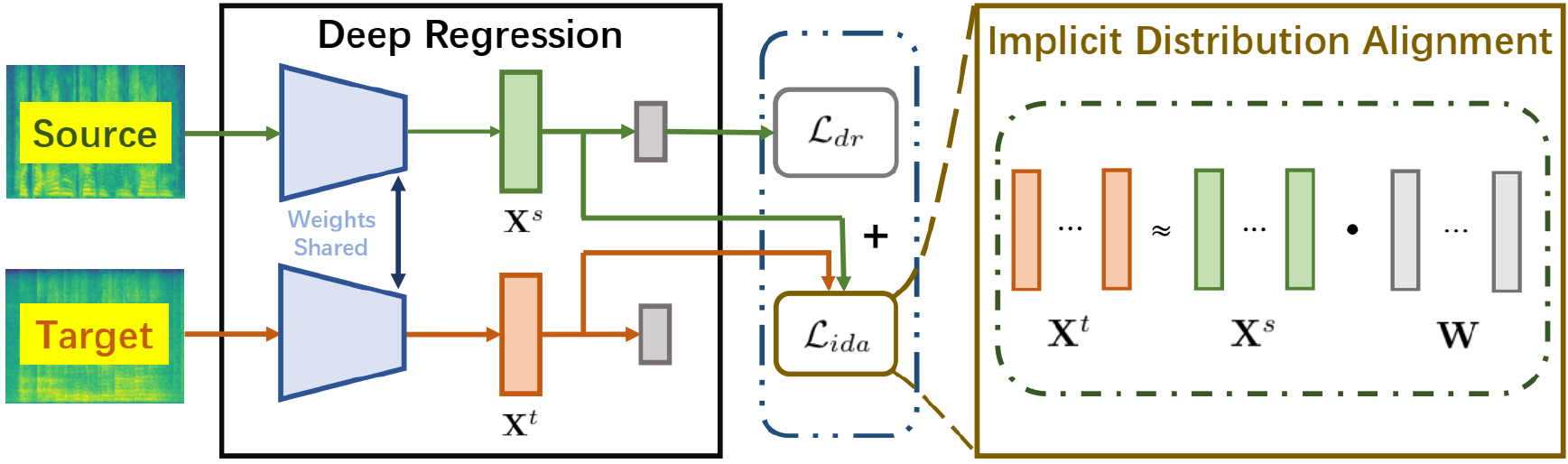}
\caption{Overview Structure and Basic Idea of the Proposed DIDAN for Dealing with the Cross-Corpus SER Problem.}
\label{Fig1}
\end{figure}

In this section, we will address the proposed DIDAN for dealing with the cross-corpus SER problem in detail. Suppose we are given a source speech emotion corpus whose samples are denoted by $\mathcal D_{s} = \{(\mathcal{X}_{i}^{s}, \mathbf y_{i}^{s})\}_{i=1}^{N_s}$, where $\mathcal{X}_{i}^{s}$ and $\mathbf y_{i}^{s}$ are the $i^{th}$ source speech spectrum and its corresponding one-hot emotion class label, and $N_s$ is the source sample number, respectively. Similarly, the unlabeled speech samples from the target corpus can be denoted by $\mathcal D_{t} = \{(\mathcal{X}_{i}^{t}, \mathbf y_{i}^{t})\}_{i=1}^{N_t}$, where $\mathcal{X}_{i}^{t}$ and $N_t$ represent the $i^{th}$ target speech spectrum and the target sample number, respectively. To make the readers better understand the proposed DIDAN, we draw an overall picture shown in Fig.~\ref{Fig1} to illustrate the basic idea and network structure. As Fig.~\ref{Fig1} shows, our DIDAN has two major parts including \textbf{Deep~Regression} and \textbf{Implicit~Distribution~Alignment}. In what follows, we will describe them in sequence.

\subsection{Deep Regression}

In our DIDAN, we first build a simple deep regression consisting of a set of convolutional and fully connected layers to bridge the source speech spectrums and their corresponding emotion labels to own the emotion discriminative ability. To achieve this goal, we can optimize the deep regression loss as follows:
\begin{equation}
\mathcal{L}_{dr}=\frac{1}{N_{s}} \sum_{i=1}^{N_{s}} J(g(f(\mathcal{X}_i^s)), \mathbf{y}_{i}^{s}),
\label{eq1}
\end{equation}
where $J(\cdot)$, $f$ and $g$ denote the cross-entropy loss function, convolution and full connection operations, respectively. It is clear to see that by feeding the source speech samples to the deep regression network and minimizing the above loss function, the DIDAN can gradually be aware of how to distinguish different emotional speech signals.

\subsection{Implicit Distribution Alignment}

Subsequently, we design a novel loss function called IDA to enable DIDAN to be also applicable to recognizing the emotions of target speech signals. Specifically, instead of measuring and narrowing the feature distribution gap like MMD based methods, we would like to make each target speech feature learned by DIDAN possibly look like a source one. To this end, together with deep regression loss function, the following regularization term should be also included in optimization of DIDAN, which can be expressed as:
\begin{equation}
\mathcal{L}_{ida} = \Vert \mathbf{X}^t - \mathbf{X}^s \mathbf{W} \Vert^{2}_{F} + \alpha \Vert \mathbf{W}\Vert_{1},
\label{eq2}
\end{equation}
where $\mathbf{X}^s = [f(\mathcal{X}_1^s), \cdots, f(\mathcal{X}_{N_s}^s)] \in \mathbb{R}^{d\times N_s}$, $\mathbf{X}^t = [f(\mathcal{X}_1^t), \cdots, f(\mathcal{X}_{N_t}^t)] \in \mathbb{R}^{d\times N_t}$, $\mathbf{W} = [\mathbf{w}_{1},\cdots,\mathbf{w}_{N_t}] \in \mathbb{R}^{N_s \times N_t}$ is a reconstruction coefficient matrix whose $i^{th}$ column corresponds to the $i^{th}$ target speech sample, and $\alpha$ is a trade-off parameter. It is also noted that $\Vert \mathbf{W}\Vert_{1} = \sum_{i=1}^{N_t}\Vert \mathbf{w}_i\Vert_1$ is a $L_1$ norm with respect to the reconstruction coefficient matrix. By minimizing such norm, DIDAN would produce a sparse $\mathbf{w}_i$, which means only a few source samples are needed to reconstruct $i^{th}$ target one.

\subsection{Total Loss Function}

By combining Eqs.(\ref{eq1}) and (\ref{eq2}), we will arrive at the final total loss function for learning DIDAN, whose corresponding optimization problem is as follows:
\begin{equation}
\min_{\theta_{f}, \theta_{g}, \mathbf{W}} \mathcal{L}_{dr}+\lambda \mathcal{L}_{ida},
\label{eq1}
\end{equation}
where $\theta_{f}$ and $\theta_{g}$ denote the network parameters corresponding to the convolutional operation $f$ and full connection operation $g$, respectively, and $\lambda$ is the trade-off parameter balancing the deep regression and IDA losses.

\section{Experiments}

\subsection{Speech Emotion Corpora and Experimental Setup}
To evaluate the proposed DIDAN, three public available speech emotion corpora, \emph{i.e.}, EmoDB (B)~\cite{2005A}, eNTERFACE (E)~\cite{DBLP:conf/icde/MartinKMP06}, and CASIA (C)~\cite{zhang2008design}, are employed to design the cross-corpus SER experiments. EmoDB is a German speech emotion corpus consisting of 535 speech samples from 10 speakers in total. These speakers were requested to perform the seven pre-defined emotional contexts including \textit{Happy} (HA), \textit{Sad} (SA), \textit{Disgust} (DI), \textit{Angry} (AN), \textit{Fear} (FE), \textit{Neutral} (NE), and \textit{Boredom}. Different from EmoDB, eNTERFACE is an English audio-visual bimodal emotion database and hence in the experiments we only adopt its audio data. It is collected from 43 individuals resulting 1582 samples, each of which is labeled as one of six basic emotions including HA, SA, FE, AN, \textit{Surprise} (SU) and \textit{Disgust} (DI). As for CASIA, it is a Chinese speech emotion corpus. It has four different speakers and totally 1200 speech samples. Each speaker is required to perform utterances with six different emotions, \emph{i.e.}, AN, SA, FE, HA, NE, and SU, respectively. 

In the task of cross-corpus SER, one speech corpus is served as the source one and the other different corpus as the target one. Therefore, by alternatively choosing either two of the above three speech emotion corpora and meanwhile extracting the speech samples sharing the same emotion labels, we are able to obtain six cross-corpus SER tasks, which can be denoted by B $\rightarrow $ E, E $\rightarrow$ B, B $\rightarrow$ C, C $\rightarrow$ B, E $\rightarrow$ C, and C $\rightarrow$ E, respectively. Note that the right and left corpora of the arrow correspond to the source and target ones. Table~\ref{tab:nb1} summarizes the statistical information of speech samples associated with these six tasks. Moreover, following the pioneer work in cross-corpus SER~\cite{schuller2010cross}, we adopt unweighted average recall (UAR), which is defined as the average of the prediction accuracy per class, to serve as the performance metric.

\begin{table}[t!]
\footnotesize
\centering
\renewcommand{\arraystretch}{0.9}
\caption{The sample statistics of corpora used in the designed six cross-corpus SER tasks.}
\begin{tabular}{|c|c|c|}
\hline
\textbf{Tasks} & \textbf{Speech Corpus (\# Samples of Each Emotion)} & \textbf{Total} \\ \hline\hline
B $\rightarrow$ E & B (AN: 127, SA: 62, FE: 69, HA: 71, DI: 46) & 375 \\
E $\rightarrow$ B & E (AN: 211, SA: 211, FE: 211, HA: 208, DI: 211) & 1052 \\\hline
B $\rightarrow$ C & B (AN: 127, SA: 62, FE: 69, HA: 71, NE: 79) & 408 \\
C $\rightarrow$ B & C (AN: 200, SA: 200, FE: 200, HA: 200, NE: 200) & 1000 \\ \hline
E $\rightarrow$ C & E (AN: 211, SA: 211, FE: 211, HA: 208, SU: 211) & 1052 \\
C $\rightarrow$ E & C (AN: 200, SA: 200, FE: 200, HA: 200, SU: 200) & 1000 \\ \hline
\end{tabular}
\label{tab:nb1}
\end{table}

\begin{table*}[t!]
\centering
\small 
\renewcommand{\arraystretch}{0.95}
\caption{The results of all transfer learning methods for cross-corpus SER tasks, where the best results are highlighted in bold.}
\begin{tabular}{|c|c|cccccc|c|}
\hline
\multicolumn{2}{|c|}{\textbf{Method}} & \textbf{B$\rightarrow$ E} & \textbf{E$\rightarrow$B} & \textbf{B$\rightarrow$C} & \textbf{C$\rightarrow$B} & \textbf{E$\rightarrow$C} & \textbf{C$\rightarrow$E} & \textbf{Average}\\
\hline \hline
\multirow{8}{*}{Subspace Learning} & SVM & 28.9 & 23.6 & 29.6 & 35.0 & 26.1 & 25.1 & 28.1\\
& TCA & 30.5 & 44.0 & 33.4 & 45.1 & 31.1 & 32.3 & 36.1\\
& GFK & 32.1 & 42.5 & 33.1 & 48.1 & \textbf{32.8} & 28.1 & 36.1\\
& SA   & 33.5 & 43.9 & 35.8 & 44.0 & 32.6 & 28.2 & 36.3\\
& DoSL & 36.1 & 39.0 & 34.4 & 45.8 & 30.4 & 31.6 & 36.2\\
& JDAR & 36.3 & 40.0 & 31.1 & 46.3 & 32.4 & 31.5 & 36.3 \\
& JIASL &\textbf{36.9} & 44.1 & 36.5 & 49.3 & 30.5 & \textbf{33.2} & 38.4\\
 \hline
\multirow{7}{*}{Deep Learning} & VGG-11 & 32.8 & 38.8 & 36.4 & 50.0 & 27.1 & 30.0 & 35.9 \\
& DAN & 35.2 & 39.2 & 36.7 & 51.6 & 28.5 & 32.5 & 37.3\\
& JAN & 34.9 & 39.6 & 37.4 & 52.1 & 27.9 & 29.8 & 37.0\\
& DANN & 35.0 & 43.6 & 37.6 & 52.3 & 28.9 & 30.0 & 37.9\\
& CDAN & 32.9 & 40.9 & 37.9 & 49.5 & 30.7 & 30.5 & 37.1\\
& DSAN & 35.6 & 44.0 & 38.5 & 53.4 & 30.3 & 31.7 & 38.9 \\\cline{2-9}
& DIDAN (Ours) & 36.1 & \textbf{46.0} & \textbf{39.1} & \textbf{54.5} & 31.9 & 30.9 & \textbf{39.8}\\\hline
\end{tabular}
\label{tab:nb2}
\end{table*}

\subsection{Comparison Methods and Implementation Detail}

In order to evaluate the effectiveness of the proposed DIDAN, several state-of-the-art transfer subspace learning and deep transfer learning methods are chosen to conduct the comparison experiments. The transfer subspace learning methods include transfer component analysis (TCA)~\cite{pan2010domain}, geodesic flow kernel (GFK)~\cite{gong2012geodesic}, subspace alignment (SA)~\cite{fernando2013unsupervised}, domain-adaptive subspace learning (DoSL)~\cite{liu2018unsupervised}, joint distribution adaptive regression (JDAR)~\cite{zhang2021cross} and  joint distribution implicitly aligned subspace learning (JIASL) \cite{lu2022implicitly}. Note that for these subspace learning methods, we use openSMILE toolkit~\cite{eyben2010opensmile} to extract IS09 speech feature sets~\cite{schuller2009interspeech}, which consists of 32 low-level acoustic descriptors and 12 statistical functions, to describe the speech signals in all three speech corpora. In the experiments, the elements in IS09 feature set are normalized between 0 and 1. As for the deep learning ones, deep adaptation network (DAN)~\cite{long2015learning}, domain-adversarial neural network (DANN)~\cite{ajakan2014domain}, and conditional domain adversarial network (CDAN)~\cite{long2018conditional}, and deep subdomain adaptation network (DSAN)~\cite{zhu2020deep} are adopted. Unlike the subspace learning methods, the inputs of the deep neural networks are the speech spectrums converted by applying Fourier Transformation to the original speech signals instead of the hand-crafted IS09 feature set.

In the experiments, we follow existing cross-corpus SER works by searching the hyper-parameters for all the comparison methods from a preset parameter interval to report their best results. Specifically, for TCA, GFK, and SA, we search its reduced feature dimension from $[5:5:d_{max}]$. As for DoSL and JDAR, both of them have two trade-off parameters, \emph{i.e.}, $\lambda$ controlling the distribution alignment term and $\mu$ corresponding to sparsity of projection matrix, whose searching interval is set as $[5:5:200]$. For all the deep learning methods, VGG-11 is adopted to serve as the backbone and hence the speech spectrums are resized to $224\times224$ pixels. We also include the original VGG-11 in the comparison. We search the trade-off parameters of deep learning methods from the hyper-parameter set $\{0.1:0.1:1,5,10,50,100\}$. The mini-batch stochastic gradient descent strategy is used for learning the optimal parameters of the deep learning methods. The batch sizes of the source and target speech samples are both fixed at 32.

\begin{table}[t]
\centering
\footnotesize
\caption{The ablation analysis for the proposed DIDAN.}
\renewcommand{\arraystretch}{0.92}
\begin{tabular}{|l|ccc|}
\hline
\textbf{Method} & B $\rightarrow$ E & E $\rightarrow$ B& B $\rightarrow$ C\\
\hline\hline
VGG-11 (DIDAN \emph{w/o} IDA) & 32.8 & 38.8 & 36.4\\
DIDAN \emph{w} nonSR-IDA& 34.5 & 42.9 & 37.7\\
\textbf{DIDAN \emph{w} IDA} & \textbf{36.1} & \textbf{46.0} & \textbf{39.1} \\
\hline
DAN (MMD) & 35.2 & 39.2 & 36.7\\
DANN (Adversarial Learning) & 35.0 & 43.6 & 37.7 \\ \hline
\end{tabular}
\label{tab:nb3}
\end{table}

\subsection{Results and Discussions}

The detailed experimental results are given in Table~\ref{tab:nb2}. Several interesting observations and conclusions can be obtained. 

(1) It is clear to see that the proposed DIDAN method achieved the best average UAR reaching 39.8\% among all the methods. Moreover, we also observed that our DIDAN outperformed all the comparison methods including subspace learning and deep learning ones in three of all six cross-corpus SER tasks, \emph{i.e.}, E $\rightarrow$ B (46.0\%), B $\rightarrow$ C (39.1\%), and C $\rightarrow$ B (54.5\%). Nevertheless, it can be found that the performance of our DIDAN is actually very competitive against the best-performing ones in the rest three tasks, \emph{i.e.}, 36.1\% (DIDAN) \emph{v.s.} 36.9\% (JIASL) in B $\rightarrow$ E, 31.9\% (DIDAN) \emph{v.s.} 32.8\% (GFK) in E $\rightarrow$ C, and 30.9\% (DIDAN) \emph{v.s.} 33.2\% (JIASL) in C $\rightarrow$ E. The above observations demonstrated the effectiveness and superior performance of the proposed DIDAN in dealing with the problem of cross-corpus SER. 

(2) Overall speaking, the deep learning methods showed more promising performance in dealing with cross-corpus SER tasks than the subspace learning ones, which can be seen from the comparison between their average UAR. Despite of this, it is noticed that some subspace learning methods can still obtain more satisfactory results compared with the deep learning ones when coping with several cross-corpus SER tasks, \emph{e.g.}, JIASL (36.9\%) in B $\rightarrow$ E and GFK (32.8\%) in E $\rightarrow$ C. This may be because that the hand-crafted speech feature set used in subspace learning methods, \emph{i.e.}, IS09, is more discriminative and corpus invariant than the deep features directly learned from the speech spectrums by the backbone (VGG-11) used in deep learning ones for these tasks. By further comparing the results of all the methods in these tasks, it is clear that the deep learning methods mostly performed poorer compared with the subspace learning ones, which supports our explanations and analysis. 

(3) It can be observed that nearly all the methods cannot well cope with the cross-corpus SER tasks between eNTERFACE and CASIA. We believe that this may attribute to the large differences between these two speech corpora. According to the works of~\cite{DBLP:conf/icde/MartinKMP06,zhang2008design}, it is well known that eNTERFACE is an English speech corpus whose samples are elicited by well-designed induced paradigm, while CASIA is a Chinese one, in which all speakers are requested to simulate different emotions.


\subsection{Going Deeper into IDA of DIDAN}

As described previously, one of the major contributions in our DIDAN is the IDA loss, which aims to improve the robustness of DIDAN to the corpus variance by enforcing the target speech features to be sparsely reconstructed by the learned source ones. To see whether IDA indeed works, we select three cross-corpus SER tasks, \emph{i.e.}, B $\rightarrow$ E, E $\rightarrow$ B, and B $\rightarrow$ C, as representatives to conduct additional experiments. Besides the original DIDAN (denoted by DIDAN \emph{w} IDA), another four methods are chosen including VGG-11 (DIDAN \emph{w/o} IDA), DIDAN without sparsity regularization term (denoted by DIDAN \emph{w} nonSR-IDA), DAN (MMD), and DANN (Adversarial Learning). Experimental results are depicted in Table~\ref{tab:nb3}. From the results in first three rows of Table~\ref{tab:nb3}, it can be concluded that the proposed implicit distribution alignment method can remarkably improve the corpus invariant ability of the deep regression model. In addition, we can also observe that our DIDAN outperformed DAN and DANN, which adopt MMD and GAN to align the feature distribution gap, respectively. This verifies the superiority of the proposed IDA in DIDAN over these two widely-used strategies for distribution alignment.

\section{Conclusion}

In this paper, we have presented a novel deep transfer learning method called DIDAN for dealing with the problem of cross-corpus SER. Unlike most of existing deep learning methods, our DIDAN removes the feature distribution mismatch between the source and target speech signals with an implicit manner, \emph{i.e.}, enforcing the target deep features to be sparsely reconstructed by the source ones. Hence, the deep regression model only supervised by the source label information would be able to effectively recognize the emotions of unlabeled target speech signals. Extensive experiments were carried out on three widely-used speech emotion corpora to evaluate the performance of the proposed DIDAN. The results showed that compared with recent state-of-the-art transfer subspace learning and deep transfer learning methods, our DIDAN has more promising performance in dealing with cross-corpus SER tasks.

\section{Acknowledgements}
This work was supported in part by the National Key R \& D Project under the Grant 2022YFC2405600, in part by the NSFC under the Grants U2003207 and 61921004, in part by the Jiangsu Frontier Technology Basic Research Project under the Grant BK20192004, and in part by the Zhishan Young Scholarship of Southeast University.
\bibliographystyle{IEEEbib}
\small
\bibliography{ICASSP2022}

\end{document}